\documentstyle[preprint,aps]{revtex}

\begin{document}
\title{M\"{o}ssbauer Effect Probe of Local Jahn-Teller distortion in Fe-doped
Colossal Magnetoresistive Manganites}
\author{Zhao-hua Cheng$^{\ast }$, Zhi-hong Wang, Nai-li Di, Zhi-qi Kou, Guang-jun
Wang, Rui-wei Li, Yi Lu, Qing-an Li, and Bao-gen Shen}
\address{State Key Laboratory of Magnetism and International Center for Quantum\\
Structures, Institute of Physics, Chinese Academy of Sciences, Beijing\\
100080, P.R. China}
\author{R.A. Dunlap}
\address{Department of Physics, Dalhousie University, Halifax, NovaScotia,Canada B3H\\
3J5 }
\maketitle

\begin{abstract}
Local structure of the Fe-doped La$_{1-x}$Ca$_{x}$MnO$_{3}$ (x=0.00-1.00)
compounds has been investigated by means of M\"{o}ssbauer spectroscopy. $%
^{57}$Fe M\"{o}ssbauer spectra provide a direct evidence of Jahn-Teller
distortion in these manganites. On the basis of M\"{o}ssbauer results, the
Jahn-Teller coupling was estimated. It is noteworthy that Ca-concentration
dependence of Jahn-Teller coupling strength is very consistent with the
magnetic phase diagram. Our results reveal that M\"{o}ssbauer spectroscopy
can not only detect the local structural distortion, but also provide a
technique to investigate Jahn-Teller coupling of Fe-doped La$_{1-x}$Ca$_{x}$%
MnO$_{3}$ colossal magnetoresistive perovskites.


*Corresponding author
\end{abstract}


The discovery of colossal magnetoresistance (CMR) in the manganese
perovskites La$_{1-x}$Ca$_{x}$MnO$_{3}$ has recently attracted significant
scientific attention.$^{1,2}$ More and more evidence has indicated that the
double-exchange coupling model alone is not sufficient to explain the CMR
effect, and strong electron-lattice coupling plays an important role in
determining the observed resistivity behavior and the magnetic transition
temperature.$^{3-7}$ Millis has argued that there are two types of lattice
effects in ABO$_{3}$-type magnetoresistive manganese perovskites.$^{7}$ One
is the tolerance factor, $t=\frac{R_{A}+R_{O}}{\sqrt{2}(R_{B}+R_{O})}$,
where $R_{A}$, $R_{B}$ and $R_{O}$ are the radii of A, B and O ions,
respectively.$^{4}$ The other one is the local structural distortion of MnO$%
_{6}$ octahedron resulting from the Mn$^{3+}$ Jahn-Teller effect. Since the
long-range structural distortion decreases rapidly in La$_{1-x}$Ca$_{x}$MnO$%
_{3}$ compounds with increasing Ca concentration,$^{8}$ a local structural
probe, such as M\"{o}ssbauer spectroscopy is required to analyze the
distortion of the MnO$_{6}$ octahedron.

The idea of using M\"{o}ssbauer spectroscopy to detect local structure is
based on the quadrupole splitting ($\Delta $) that is very sensitive to the
distribution of surrounding electrons of resonant nuclei. A few percent of
Fe substitution for Mn in La$_{1-x}$Ca$_{x}$MnO$_{3}$ compounds can be used
as a micro-probe to detect the symmetry of the nearest-neighbor O$^{2-}$
ions in the Mn(Fe)O$_{6}$ octahedron. Furthermore, early studies have shown
that Mn$^{3+}$ ions are mainly replaced by Fe$^{3+}$ ions in this Fe-doping
range,$^{9}$ and that both ions have identical ionic radii (0.645 \r{A}) in
six-fold octahedral coordination.$^{10}$ Therefore, the substitution of Fe$%
^{3+}$ for Mn$^{3+}$ does not change the tolerance factor, and consequently,
the Jahn-Teller effect can be investigated separately.

Several papers have been published using M\"{o}ssbauer spectroscopy,$%
^{11-14} $ but the local structural distortion of Mn(Fe)O$_{6}$ octahedra
and Jahn-Teller effect were not yet investigated by this technique. In this
Letter, we focused our investigation on the local structural information
regarding Mn(Fe)O$_{6}$ octahedra. We present $^{57}$Fe M\"{o}ssbauer
spectra recorded by a Wissel System constant acceleration M\"{o}ssbauer
spectrometer with a $^{57}$Co(Pd) source that not only clearly indicate the
presence of quadrupole splitting but also show the occurrence of a second
quadrupole split doublet. As an example, M\"{o}ssbauer spectrum of La$%
_{0.69} $Ca$_{0.31}$Mn$_{0.91}$Fe$_{0.09}$O$_{3}$ polycrystalline powders
prepared by conventional solid-state reaction method illustrated in Fig.
1(a) can be best fitted with two doublets. The more intense doublet ($\simeq 
$90\%) has a center shift $\delta $ = 0.339$\pm $0.010 mm/s relative to $%
\alpha $-Fe at room temperature, and a quadrupole splitting $\Delta $= 0.235$%
\pm $0.021 mm/s, which are quite similar with previous reports.$^{11-14}$
This center shift is typical value of high-spin Fe$^{3+}$ with octahedral
coordination. The weaker doublet ($\simeq $10\%) has $\delta $ = 0.147$\pm $%
0.043 mm/s and $\Delta $ = 0.672$\pm $0.085 mm/s. This center shift is in
good agreement with that of low-spin Fe$^{4+}$ in SrFeO$_{3}$ compound .$%
^{15}$ The quadrupole splitting for both Fe$^{3+}$ and Fe$^{4+}$ ions
confirms the local distortion of the Mn(Fe)O$_{6}$ octahedron.

Since too much Fe as a dopant alters the electronic and magnetic properties
significantly,$^{12,16}$ we selected some typical samples (x=0.31, 0.50 and
0.60) containing smaller concentration of Fe(4at.\%) to reduce the Fe-O-Fe
contact and carried out x-ray diffraction (XRD) and M\"{o}ssbauer
measurements again. XRD pattern indicated that there is no change in
structure and lattice parameters between samples with different Fe
concentrations due to the same radii of Mn$^{3+}$ and Fe$^{3+}$ ions.
M\"{o}ssbauer spectra of samples with lower Fe concentration demonstrate
that the center shift $\delta $, quadrupole splitting $\Delta $ and line
width corresponding to Fe$^{3+}$ ions are in good agreement with that of Fe$%
^{3+}$ ions in the samples with 9at.\% Fe concentrations. These measurements
confirm that Fe-O-Fe contacts have no obvious contribution to M\"{o}ssbauer
parameters at temperatures T%
\mbox{$>$}%
T$_{C}$. The doublet originated from low-spin Fe$^{4+}$ ion was not detected
due to the lower Fe substitution. M\"{o}ssbauer spectra of the samples
containing lower Fe concentration also indicate a local distortion of the
Mn(Fe)O$_{6}$ octahedron in La-Ca-Mn-O perovskites.The fact that the
Jahn-Teller distortion in the perovskites with x 
\mbox{$>$}%
0.2 is detected by M\"{o}ssbauer effect, rather than by XRD technique
implies that this distortion is a dynamic one, rather than a static one
since the time scale probed by M\"{o}ssbauer effect is much smaller than
that of XRD technique. One possible reason for the different information
obtained by these two techniques is due to their different time scales.

The local structural distortion of MnO$_{6}$ octahedra resulting from the
Jahn-Teller effect of high-spin Mn$^{3+}$ ions removes the degeneracy of the 
$e_{g}$ and $t_{2g}$ orbitals so as to make some energy levels more stable.
The $e_{g}$ orbital group is separated into two energy levels, $d_{z^{2}}$
and $d_{x^{2}-y^{2}}$. The $t_{_{2g}}$ orbital group is split into a
non-degenerate energy level and a two-fold degenerate energy level in the
trigonally or tetragonally distorted octahedra. The two-fold degenerate
energy level will further split in highly distorted octahedra. The energy
separation of the upper-level orbitals has been shown to be larger than that
of the lower-level orbitals. Since the Jahn-Teller distortion strongly
influences the electron hopping process between the upper-level orbitals of
Mn$^{3+}$ and Mn$^{4+}$ ions in La$_{1-x}$Ca$_{x}$MnO$_{3}$ perovskites, and
consequently determines the ferromagnetic and electrical behaviors, we only
need to consider the upper-level splitting of $e_{g}$ orbitals. The energy
separation between $d_{z^{2}}$ and $d_{x^{2}-y^{2}}$, $E_{JT}$, can be
described as$^{17}$

$E_{JT}=4D_{s}+5D_{t}$ \ \ \ \ \ \ \ \ \ \ \ \ \ \ \ \ \ \ \ \ \ \ \ \ \ \ \
\ \ \ \ \ \ \ \ \ \ \ \ \ \ \ \ \ \ \ \ \ \ \ \ \ \ \ \ \ \ \ \ \ \ \ \ \ \
\ \ \ \ \ \ \ \ \ \ \ \ \ \ \ \ \ \ \ \ \ \ (1)

$D_{s}=\frac{e}{14}\sqrt{\frac{5}{\pi }}A_{20}<r^{2}>_{3d}\ \ $\ \ \ \ \ \ \
\ \ \ \ \ \ \ \ \ \ \ \ \ \ \ \ \ \ \ \ \ \ \ \ \ \ \ \ \ \ \ \ \ \ \ \ \ \
\ \ \ \ \ \ \ \ \ \ \ \ \ \ \ \ \ \ \ \ \ \ \ \ \ (2)

$D_{t}=\frac{e}{14\sqrt{\pi }}T<r^{4}>_{3d\ }$\ \ \ \ \ \ \ \ \ \ \ \ \ \ \
\ \ \ \ \ \ \ \ \ \ \ \ \ \ \ \ \ \ \ \ \ \ \ \ \ \ \ \ \ \ \ \ \ \ \ \ \ \
\ \ \ \ \ \ \ \ \ \ \ \ \ \ \ \ \ \ \ \ \ \ \ \ \ (3)

where $A_{20}$=$\sqrt{\frac{\pi }{5}}\sum \frac{q_{i}(3\cos ^{2}\theta
_{i}-1)}{R_{i}^{3}}$ is the second-order crystal-electric-field (CEF)
coefficient. The electric charge, $q_{_{i}}$, of the i$^{th}$ surrounding
ion should be effective charge as determined by a consideration of the
shielding and polarization of inner-shell electrons of the central ion. $%
R_{i}$ is the radial distance of the ith surrounding ion from the central
ion, and $\theta _{i}$ is its polar angle in spherical coordinates. $A_{20}$
is zero for cubic or perfectly octahedral symmetry. $T=A_{40}(oct^{\prime
})-A_{40}(oct)$ is the difference between the fourth-order CEF coefficient
of a distorted MnO$_{6}$ octahedron, $A_{40}(oct^{\prime })$ , and that for
perfect octahedral symmetry, $A_{40}(oct)$, where $A_{40}=\frac{\sqrt{\pi }}{%
4}\sum \frac{q_{i}}{R_{i}^{5}}(\frac{35}{3}\cos ^{4}\theta _{i}-10\cos
^{2}\theta _{i}+1)$. Finally, 
\mbox{$<$}%
$r^{n}$%
\mbox{$>$}%
$_{3d}$ is the expectation value of the n$^{th}$ power, $r^{n}$ , of the
radial distance of a 3$d$ orbital from the nucleus.

Using the Mn-O bond lengths and the expectation values of 
\mbox{$<$}%
$r^{2}$%
\mbox{$>$}%
$_{3d}$ and 
\mbox{$<$}%
$r^{4}$%
\mbox{$>$}%
$_{3d}$ for Mn$^{3+}$ ions,$^{18}$ it can easily be shown from eqs. (2) and
(3) that $D_{t}<<D_{s}.$ Therefore, the energy separation between and
orbitals arises mainly from the contribution of the second-order CEF
coefficient $A_{20}$ of the distorted octahedra and eq. (1) can be written
as:

$E_{JT}\simeq 4Ds=\frac{2e}{7}\sqrt{\frac{5}{\pi }}A_{20}<r^{2}>_{3d}$ \ \ \
\ \ \ \ \ \ \ \ \ \ \ \ \ \ \ \ \ \ \ \ \ \ \ \ \ \ \ \ \ \ \ \ \ \ \ \ \ \
\ \ \ \ \ \ \ \ \ \ \ \ \ \ \ \ \ \ \ (4)

Although the relationship between $E_{JT}$ and $A_{20}$ is relatively
simple, small errors in X-ray structural data can cause very large errors in
the values of $A_{20}$. Moreover, since the surrounding ionic charge is not
concentrated exclusively at the lattice positions as assumed in the point
charge model, the contribution of inner-shell electrons' shielding and the
polarization of the central ions should be taken into account as well.
Therefore, it is difficult to obtain a reliable result on the basis of x-ray
diffraction data alone. However, M\"{o}ssbauer spectroscopy can be employed
to determine the value of $A_{20}$ by a measurement of the quadrupole
splitting as described below.

In the case of $^{57}$Fe M\"{o}ssbauer effect, the quadrupole splitting, $%
\Delta $, can be written as$^{19}$

$\Delta =\frac{eQV_{zz}}{2}(1+\frac{\eta ^{2}}{3})^{1/2}$ \ \ \ \ \ \ \ \ \
\ \ \ \ \ \ \ \ \ \ \ \ \ \ \ \ \ \ \ \ \ \ \ \ \ \ \ \ \ \ \ \ \ \ \ \ \ \
\ \ \ \ \ \ \ \ \ \ \ \ \ \ \ \ \ \ \ \ \ \ \ \ \ \ \ \ \ \ \ \ (5)

where $Q$ is the electric quadrupole moment of the $^{57}$Fe nucleus, $\eta $
is the asymmetry parameter and $V_{zz}$ is the principal component of the
electric field gradient (EFG) at the nucleus. This final quantity includes
contributions from valence electrons of both the Fe ion $V_{ZZ}$(Fe), and
the surrounding ions, $V_{ZZ}$(latt). This may be written as

$V_{ZZ}(tot)=V_{ZZ}(Fe)+V_{ZZ}(latt)$ \ \ \ \ \ \ \ \ \ \ \ \ \ \ \ \ \ \ \
\ \ \ \ \ \ \ \ \ \ \ \ \ \ \ \ \ \ \ \ \ \ \ \ \ \ \ \ \ \ \ \ \ \ \ \ \ \
\ \ \ \ (6)

where

$V_{ZZ}(latt)=\sum \frac{q_{i}(3\cos ^{2}\theta _{i}-1)}{R_{i}^{3}}=\sqrt{%
\frac{5}{\pi }}A_{20}$ \ \ \ \ \ \ \ \ \ \ \ \ \ \ \ \ \ \ \ \ \ \ \ \ \ \ \
\ \ \ \ \ \ \ \ \ \ \ \ \ \ \ \ \ \ \ \ \ \ \ \ \ (7)

$V_{ZZ}(Fe)=\sum P_{s}V_{zz}(s)$\ \ \ \ \ \ \ \ \ \ \ \ \ \ \ \ \ \ \ \ \ \
\ \ \ \ \ \ \ \ \ \ \ \ \ \ \ \ \ \ \ \ \ \ \ \ \ \ \ \ \ \ \ \ \ \ \ \ \ \
\ \ \ \ \ \ \ \ \ \ \ \ \ \ \ \ (8)

where $P_{s}$ is the probability of the valence electrons at different
orbitals. The values of $V_{zz}$(s) for $d_{x^{2}-y^{2}},$ $d_{z^{2}}$, $%
d_{xy}$, $d_{yz}$ and $d_{zx}$ orbitals are $+\frac{4e}{7}<r^{-3}>,-\frac{4e%
}{7}<r^{-3}>$ , $+\frac{4e}{7}<r^{-3}>$, $-\frac{2e}{7}<r^{-3}>,-\frac{2e}{7}%
<r^{-3}>$, respectively.$^{19}$ Thus, the contribution from the valence
electrons of the Fe cations to the EFG at the nucleus depends on their
configuration, i.e. the valence-state and spin-state. In the case of perfect
octahedra, the four valence electrons of a low-spin Fe$^{4+}$ ion occupy the
three-fold degenerate $t_{2g}$ orbitals $(d_{xy}$, $d_{yz}$ and $d_{zx}$ )
with equal probability in the ground state and produce no EFG, $V_{zz}$%
(Fe)=0. However, in the case of lower symmetry, the degeneracy is removed
and electrons will preferentially occupy the low-lying energy level and a
non-zero EFG will be produced in an Fe$^{4+}$ ion by its valence electrons, $%
V_{zz}$(Fe)$\neq $0. For high-spin Fe$^{3+}$, however, five valence
electrons each occupy one of the five 3$d$ orbitals. The half-filled shell
is spherically symmetric regardless of the distribution of surrounding ions
and the valence electrons make no contribution to the EFG, $V_{zz}$(Fe)=0.
Therefore, the valence electrons will make no contribution to the quadrupole
splitting for either Fe$^{3+}$ or Fe$^{4+}$ ions in perfect octahedral
symmetry, but will yield a contribution to the quadrupole splitting for Fe$%
^{4+}$ in distorted octahedra. Compared with Fe$^{3+}$ ion, about 0.4mm/s
larger in quadrupole splitting corresponding to Fe$^{4+}$ ions may be
related to valence electron contribution.

Since $\Delta $ at the nucleus of Fe$^{3+}$ ion is solely the result of
contributions from the surrounding ions, the values of $A_{20}$ can be
obtained from

$A_{20}$=$\sqrt{\frac{4\pi }{5}}\frac{\Delta (Fe^{3+})}{eQ(1+\frac{\eta 2}{3}%
)^{1/2}}$ \ \ \ \ \ \ \ \ \ \ \ \ \ \ \ \ \ \ \ \ \ \ \ \ \ \ \ \ \ \ \ \ \
\ \ \ \ \ \ \ \ \ \ \ \ \ \ \ \ \ \ \ \ \ \ \ \ \ \ \ \ \ \ \ \ \ \ \ \ \ \
\ \ \ \ \ \ \ \ \ \ \ \ \ \ \ \ \ \ \ \ \ \ \ \ \ \ \ \ \ \ \ \ (9)

Therefore, the relationship between $E_{JT}$ and the quadrupole splitting at
Fe$^{3+}$ ion $\Delta $(Fe$^{3+}$) can be written as

$E_{JT}=\frac{4\Delta (Fe^{3+})}{7Q(1+\frac{\eta ^{2}}{3})^{1/2}}%
<r^{2}>_{3d} $ \ \ \ \ \ \ \ \ \ \ \ \ \ \ \ \ \ \ \ \ \ \ \ \ \ \ \ \ \ \ \
\ \ \ \ \ \ \ \ \ \ \ \ \ \ \ \ \ \ \ \ \ \ \ \ \ \ \ \ \ \ \ \ \ \ \ \ \ \
\ \ \ \ \ \ \ \ \ \ \ \ \ \ \ \ \ \ \ \ \ (10)

where $\eta $ = 0 for axial symmetry. The quadrupole splitting for
M\"{o}ssbauer experiment is generally in unit of mm/s. Here it should be
converted to eV by a factor of $E\gamma /c$, where $E\gamma $=14.4 keV for $%
I $ = 3/2 $\rightarrow $1/2 transition of $^{57}$Fe and $c$=3$\times $10$%
^{11}$mm/s is the velocity of light.

Using a approximation of the expectation value 
\mbox{$<$}%
$r^{2}$%
\mbox{$>$}%
of free Mn$^{3+}$ ions, 
\mbox{$<$}%
$r^{2}$%
\mbox{$>$}%
=0.3535 \r{A}$^{2}$ and $Q$ = 0.28$\times $10$^{-24}$ cm$^{2}$, $^{18,20}$
the Jahn-Teller coupling, $E_{JT}$, in La$_{1-x}$Ca$_{x}$(Mn,Fe)O$_{3}$
compounds can be estimated. Fig. 2 illustrates the quadrupole splitting
corresponding to Fe$^{3+}$ ions and Jahn-Teller coupling in La$_{1-x}$Ca$%
_{x} $Mn$_{0.91}$Fe$_{0.09}$O$_{3}$ (x=0.00-1.00) and La$_{1-x}$Ca$_{x}$Mn$%
_{0.96} $Fe$_{0.04}$O$_{3}$ (x=0.31-0.60) as a function of Ca concentration.
It is noteworthy that the Ca-concentration dependence of the Jahn-Teller
coupling strength is consistent well with the magnetic phase diagram,
implying a direct relationship between Jahn-Teller effect as well as
electric and magnetic properties in these perovskites. The Jahn-Teller
coupling is indeed strong in LaMn$_{0.91}$Fe$_{0.09}$O$_{3}$. With
increasing Ca concentration, the energy separation between $d_{z^{2}}$ and $%
d_{x^{2}-y^{2}}$ decreases from about 1.4 eV for x=0 to about 0.83 eV for
x=0.27-0.39. The transport properties of these perovskites are governed by
the interplay of the double-exchange coupling between Mn$^{3+}$ and Mn$^{4+}$
ions as well as the Jahn-Teller coupling. For the samples with x=0.27-0.39,
the relatively weak Jahn-Teller coupling delocalizes the electrons and makes
the hoping process easier via double-exchange coupling. Therefore, these
samples show a ferromagnetic metallic state at T%
\mbox{$<$}%
T$_{C}$. In the region of 0.50$<$x$<$0.80, M\"{o}ssbauer results indicate
that the Jahn-Teller coupling becomes stronger again, which accompanied by
larger change in Mn-O bond length.$^{7,21}$ The larger energy separation
between $d_{z^{2}}$ and $d_{x^{2}-y^{2}}$ tends to trap electrons in the
low-lying energy level orbitals, and hence localize the conduction
electrons. Therefore, for the samples with x%
\mbox{$>$}%
0.5, the ground state becomes insulating and antiferromagnetic again. For
the samples with x$>$0.90, the local structural distortion decreases due to
the lack of Mn$^{3+}$ Jahn-Teller ions.

In the case of ferromagnetic materials, electric quadrupole interaction is a
perturbation term in comparison with magnetic hyperfine interaction. It will
become the main term at T%
\mbox{$>$}%
T$_{c}$. M\"{o}ssbauer spectra collected at T%
\mbox{$>$}%
T$_{C}$ provide a possibility to determine quadrupole splitting precisely.
As illustrated in fig. 3, the quadrupole splitting at Fe$^{3+}$ ion is found
to have no obvious change at T%
\mbox{$>$}%
T$_{C}$, which is same as that in La$_{1-x}$Ca$_{x}$Mn$_{0.9575}$Fe$%
_{0.0425} $O$_{3}$.$^{13}$ These results suggest no abrupt change in the
difference of Mn-O bond lengths above T$_{C}$, in agreement with the results
of extended X-ray absorption fine structure (EXAFS) and neutron pair
distribution function (PDF).$^{6,7,22}$ Further work is needed to
investigate quadrupole splitting and Jahn-Teller coupling below T$_{C}$
using $^{57}$Fe-enriched samples.

In conclusion, we employed $^{57}$Fe nucleus as a micro-probe to detect the
local structure of Mn(Fe)O$_{6}$ octahedra in Fe-doped La$_{1-x}$Ca$_{x}$MnO$%
_{3}$ pervoskites. Our work not only confirms the local structural
distortion of (Mn,Fe)O$_{6}$ octahedron, but also reveals that M\"{o}ssbauer
spectroscopy can be effectively employed to estimate the Jahn-Teller
coupling of these perovskites.

This work was supported by the State Key Project of Fundamental Research,
the National Natural Sciences Foundation of China, as well as the Natural
Sciences and Engineering Research Council of Canada. Z.H.C thanks the Killam
Foundation for financial support.

\bigskip 

\bigskip 

\bigskip 

\bigskip 

\bigskip 

\bigskip 

\bigskip 

\bigskip 

\bigskip 

\bigskip 

\bigskip 

\bigskip 

\bigskip 

\bigskip 

\bigskip 

\bigskip 

\bigskip 

\bigskip 

\bigskip 

\bigskip 

\bigskip 

\bigskip 

\bigskip 

\bigskip 

\bigskip 

\bigskip 

\bigskip 

\bigskip 

\bigskip 

\bigskip 

\bigskip 

\bigskip 

\bigskip 

\bigskip 

\bigskip 

\bigskip 

Figure captions

Figure 1. Room temperature $^{57}$Fe M\"{o}ssbauer spectra of La$_{0.69}$Ca$%
_{0.31}$Mn$_{0.91}$Fe$_{0.09}$O$_{3}$ (a) and La$_{0.69}$Ca$_{0.31}$Mn$%
_{0.96}$Fe$_{0.04}$O$_{3}$(b).

Fig. 2. Quadrupole splitting at Fe$^{3+}$ ion and Jahn-Teller coupling in La$%
_{1-x}$Ca$_{x}$Mn$_{0.91}$Fe$_{0.09}$O$_{3}$ (x=0.00-1.00) and La$_{1-x}$Ca$%
_{x}$Mn$_{0.96}$Fe$_{0.04}$O$_{3}$ (x=0.31-0.60) perovskites at room
temperature.

Fig. 3. Temperature dependence of quadrupole splitting at Fe$^{3+}$ ion and
Jahn-Teller coupling in La$_{1-x}$Ca$_{x}$Mn$_{0.91}$Fe$_{0.09}$O$_{3}$ and
La$_{1-x}$Ca$_{x}$Mn$_{0.96}$Fe$_{0.04}$O$_{3}$ (x=0.31 and 0.50)


\begin{references}
\bibitem{1}  S. Jin, T. Tiefel, M. McMormack, P.A. Fastnacht, R.Ramesh and
L.H. Chen, Science, {\bf 264}, 413 (1994).

\bibitem{2}  P.Schiffer, A.P. Ramirez, W. Bao and S.W. Cheong, Phys. Rev.
Lett. {\bf 75}, 3336(1995).

\bibitem{3}  A.J. Millis, P.B. Littlewood, and B.I. Shraiman, Phys. Rev.
Lett. {\bf 74},5144 (1995).

\bibitem{4}  H.Y.Hwang, S.W. Cheong, P.G. Radaelli, M. Marezio, and B.
Batlogg, Phys. Rev. Lett. {\bf 75}, 914 (1995).

\bibitem{5}  H. R\"{o}der, J. Zang, and A.R. Bishop, Phys. Rev. Lett. {\bf 76%
}, 1356 (1996).

\bibitem{6}  C.H. Booth, F.Bridges, G.H. Kwei, J.M. Lawrence, A.L. Cornelius
and J.J. Neumeier, Phys. Rev. Lett. {\bf 80}, 853 (1998).

\bibitem{7}  A.J. Millis, Nature, {\bf 392}, 147(1998).

\bibitem{8}  J.B.A.A. Elemans, B. Van Laar, K.R. van der Veen and B.O.
Loopstra, J. Solid State. Chem. {\bf 3}, 238 (1971).

\bibitem{9}  G.H. Jonker, Physica (Amsterdam), {\bf 20}, 1118 (1954).

\bibitem{10}  R.D. Shannon, Acta. Cryst. {\bf A32}, 751 (1976).

\bibitem{11}  A. Simopoulous, M. Pissas, G. Kallias, E. Devlin, N. Moutis,
I. Panagiotopoulous, D. Niarchos, C. Christides, and R. Sonntag, Phys. Rev.
B, {\bf 59}, 1263 (1999).

\bibitem{12}  S.B.Ogale, R.Shreekala, R.Bathe, S.K. Date, S.I. Patil, B.
Hannoyer, F.Petit and G.Marest, Phys. Rev. B., {\bf 57},7841(1998).

\bibitem{13}  B.Hannoyer,G. Marest, J.M.Greneche, R. Bathe, S.I. Patil and
S.B. Ogale, Phys. Rev. B, {\bf 61},9613(2000).

\bibitem{14}  A. Nath., J. Solid State Chem.{\bf \ 155},116(2000).

\bibitem{15}  U. Shimony and J.M. Knudsen, Phys. Rev. {\bf 144}, 361 (1966).

\bibitem{16}  K. H. Ahn, X. W. Wu, K. Liu, and C. L. Chien Phys. Rev. B {\bf %
54},15299(1996)

\bibitem{17}  J.S. Griffith, The Theory of Transition-Metal Ion (Cambridge
Univ. Press, London, 1964).

\bibitem{18}  S. Fraga, Handbook of Atomic Data (Elsevier Scientific
Publishing Company, Amsterdam, 1976).

\bibitem{19}  N.N. Greewood and T.C. Gibb, M\"{o}ssbauer Spectroscopy
(Chapman and Hall Ltd. London, 1971).

\bibitem{20}  R. R. Sharma and T. P. Das, J. Chem. Phys. {\bf 41}, 3581
(1964); J.O. Artman, J. O. Artman, A. H. Muir, Jr. and H. Wiedersich Phys.
Rev. {\bf 173}, 337(1968); R.M. Sternheimer, Phys.Rev. {\bf 130}, 1423 (1963)

\bibitem{21}  P.G. Radaelli, D.E. Cox, M. Marezio, S.W. Cheong, Phys. Rev.
B. {\bf 55},3015(1997).

\bibitem{22}  S.J.L. Billinge, R.G. DiFrancesco, G.H. Kwei, \ J.J. Neumeier
and J.D. Thompson, Phys. Rev. Lett. {\bf 77},853(1996)
\end{references}
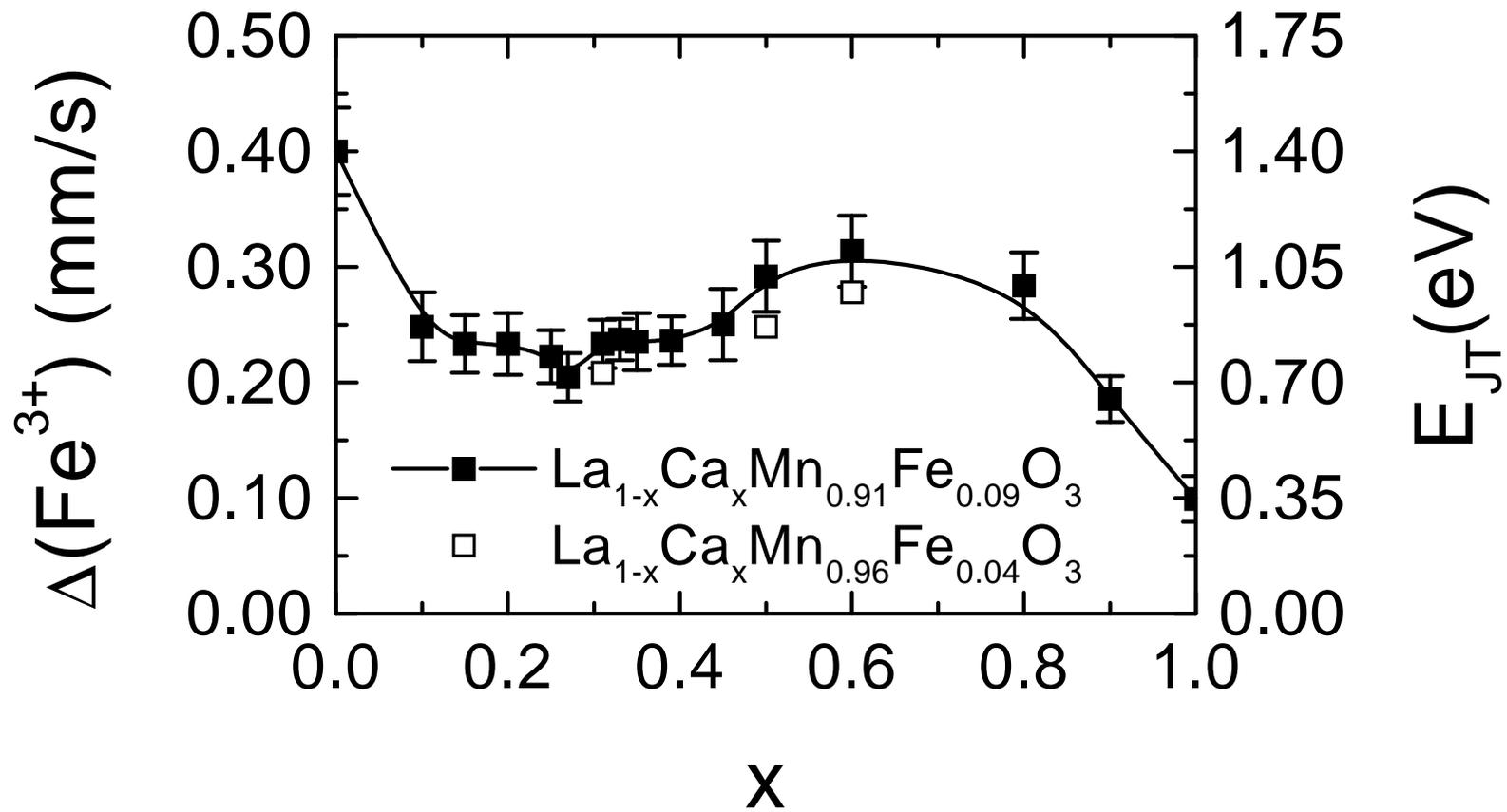
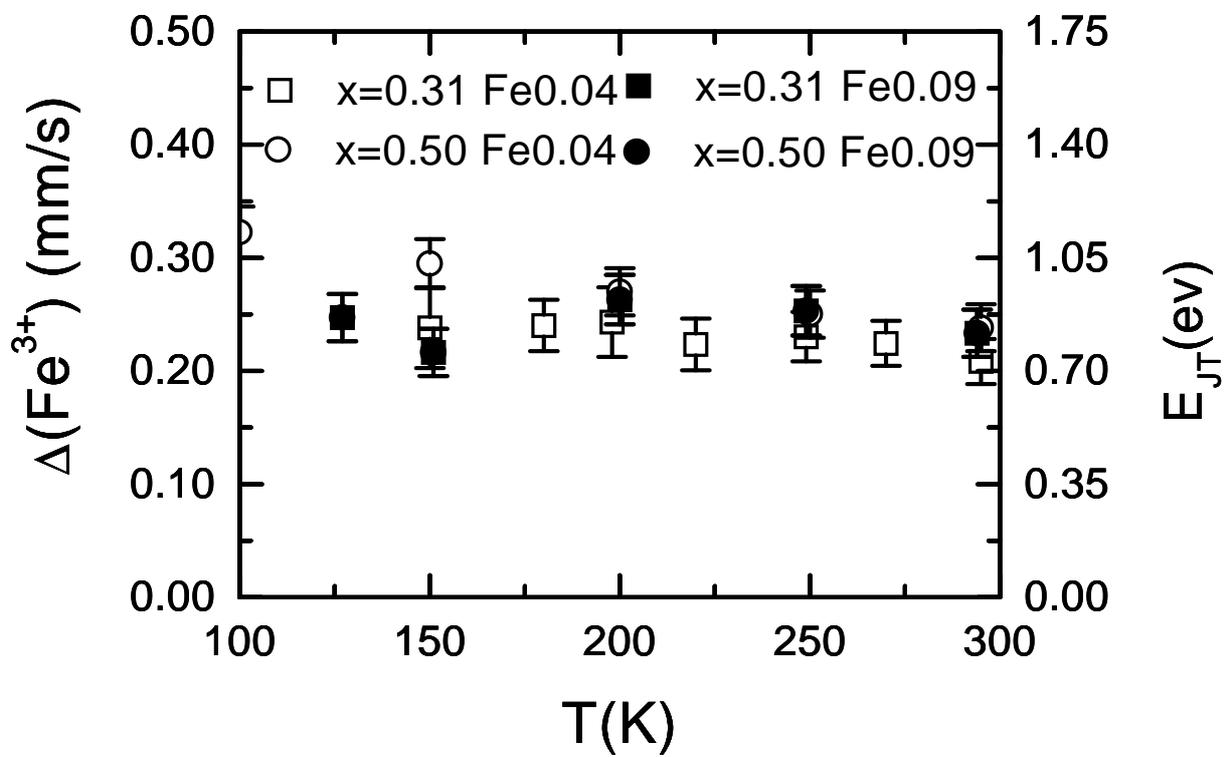
\end{document}